\documentclass[10pt,conference]{IEEEtran}
\usepackage{epsfig,rotating,setspace,latexsym,amsmath,epsf,amssymb,amsfonts,bm,theorem,cite,authblk, bbm,color, hyperref, multirow, caption, subcaption}
\IEEEoverridecommandlockouts
\allowdisplaybreaks

\title{Multi-Receiver Task-Oriented Communications via Multi-Task Deep Learning}
\begin{document}
\author[1]{Yalin E. Sagduyu}
\author[1]{Tugba Erpek}
\author[2]{Aylin Yener}
\author[3]{Sennur Ulukus}

\affil[1]{\normalsize  Virginia Tech, Arlington, VA, USA}

\affil[2]{\normalsize  The Ohio State University, Columbus, OH, USA}

\affil[3]{\normalsize University of Maryland, College Park, MD, USA}
\maketitle
\begin{abstract}
This paper studies task-oriented, otherwise known as goal-oriented, communications, in a setting where a transmitter communicates with multiple receivers, each with its own task to complete on a dataset, e.g., images, available at the transmitter. A multi-task deep learning approach that involves training a common encoder at the transmitter and individual decoders at the receivers is presented for joint optimization of completing multiple tasks and communicating with multiple receivers. By providing efficient resource allocation at the edge of 6G networks, the proposed approach allows the communications system to adapt to varying channel conditions and achieves task-specific objectives while minimizing transmission overhead. Joint training of the encoder and decoders using multi-task learning captures shared information across tasks and optimizes the communication process accordingly. By leveraging the broadcast nature of wireless communications, multi-receiver task-oriented communications (MTOC) reduces the number of transmissions required to complete tasks at different receivers. Performance evaluation conducted on the MNIST, Fashion MNIST, and CIFAR-10 datasets (with image classification considered for different tasks) demonstrates the effectiveness of MTOC in terms of classification accuracy and resource utilization compared to single-task-oriented communication systems. 
\end{abstract}


\section{Introduction} \label{sec:Intro} 
In traditional communications systems, the primary goal is to communicate messages reliably while considering impairments in the channel. To reconstruct the transmitted information from the transmitter to the receiver and minimize the loss of reconstruction, such as the symbol error rate, the operations of the transmitter and receiver can be individually or collaboratively designed. For this objective, \emph{deep neural networks} (DNNs) can serve as \emph{autoencoders}, effectively capturing both transmitter and receiver operations including channel coding and modulation at the transmitter, as well as channel decoding and demodulation at the receiver  
\cite{erpek2020deep}. The end-to-end reconstruction loss can be then effectively minimized.

The concept of \emph{semantic communications} \cite{guler2014semantic, gunduz2022beyond, uysal2021semantic} by contrast seeks to preserve the meaning of the information conveyed during communication to the receiver. To achieve this, the training loss for the autoencoder can include both the reconstruction loss from conventional communications and the semantic loss, which refers to the loss of meaning during information transfer \cite{Semanticadversarial, sagduyu2022vulnerabilities}. Semantic communications has been explored for various data types, such as text \cite{guler2018semantic, xie2021deep, kutay-yener2023icc}, image \cite{qin2021semantic, Semanticadversarial}, video \cite{Geoffreyvideo}, speech/audio \cite{weng2021semantic, walidaudio}, to maintain the integrity of meaning across different types of information, as envisioned for 6G networks.

\emph{Task-oriented communications} (TOC) or \emph{goal-oriented communications} \cite{shao2021learning, kang2022task} introduce a new approach by focusing on the semantics of information and its significance in relation to a specific task. Unlike traditional communication methods that prioritize reliable information reconstruction, the objective in TOC is to successfully accomplish a task, such as classification, using the available data at the transmitter, rather than the receiver. The transmitter's operations, including source coding, channel coding, and modulation, are modeled as an encoder that generates and transmits low-dimensional feature vectors. In TOC, the receiver deviates from the conventional receiver chain and directly employs a decoder to perform the task, such as classifying received signals, without the need for reconstructing the original input samples. TOC reduces the number of transmissions and latency for 6G applications. The encoder-decoder pair is jointly trained as an end-to-end deep neural network (DNN), considering both channel and data characteristics, to optimize task performance that can be measured with the classification loss \cite{TOCattacksagduyu} and formulated jointly with other measures such as age of information \cite{sagduyu2023age}.

Efficient utilization of limited resources is of utmost importance in wireless communication systems. As wireless networks become more complex, the need arises for communication strategies that can efficiently handle multiple tasks with multiple receivers. This paper investigates the TOC paradigm, where a single transmitter communicates with multiple receivers, each having its own task to perform on a dataset, such as images, that is originally available at the transmitter. The goal of \emph{multi-receiver task-oriented communications} (MTOC) is to jointly optimize multiple tasks and communications by employing a multi-task deep learning approach, which involves training a common encoder at the transmitter and decoders at the receivers. This approach allows for efficient resource allocation in 6G networks, enabling the adaptation to varying channel conditions, and achieves task-specific objectives while minimizing transmission overhead.

One real-world application for MTOC is \emph{image classification} in a disaster response scenario. A collaborative network of first responders can leverage their camera-enabled devices to capture and share images in real-time. Each user at the edge of 6G network has a different task related to the incident response and analysis. These tasks may include (i) damage assessment (classifying  types of damage, such as structural damage, road blockages, or hazardous materials), (ii) search and rescue (detecting signs of survivors, injured people, or potential hazards that may hinder rescue operations), (iii) resource allocation (identifying available supplies and equipment), and (iv) environmental monitoring (assessing the environmental impact of the disaster and detecting the risks, such as hazardous materials).

The key idea behind the proposed approach is to leverage the power of multi-task learning to capture shared information across tasks and optimize the communication process accordingly. By training the encoder and decoders jointly using multi-task learning, the system can efficiently allocate resources based on the specific requirements of each task. The inherent broadcast property of wireless communications is leveraged, enabling MTOC to reduce the number of transmissions required to complete tasks at different receivers.

We consider image classification to identify different tasks by using MNIST, Fashion MNIST, and CIFAR-10 datasets. We train different feedforward neural network (FNN) and convolutional neural network (CNN) models for the encoder and set of decoders at different receivers. In multi-task learning, where a single model is trained to perform multiple tasks simultaneously, losses from different tasks are combined for the training process. Each task's loss is multiplied by a task-specific weight, reflecting its relative importance. The overall loss is obtained by taking the weighted sum of the individual losses across all tasks. This way, a balance is obtained between optimizing each task individually and shared information is leveraged across tasks to improve overall performance. 

We characterize the classification accuracy for different tasks as a function of the signal-to-noise ratio (SNR) for the communication channel, the transmitter output size (reflecting the level of compression before transmitting over the channel), the weights for different tasks, and the number of receivers each with its own task to complete. The performance of MTOC is compared to that of single-task-oriented communications (STOC), where a non-zero weight is assigned only to a single task). The results demonstrate the advantages of MTOC in terms of improved classification accuracy and optimized resource utilization, showcasing its potential to enhance the performance of wireless networks in scenarios involving multiple receivers with diverse tasks.

The rest of the paper is organized as follows. Section~\ref{sec:TOC} outlines TOC, MTOC, and multi-task learning. Section~\ref{sec:data} describes the datasets and DNN architectures for MTOC. Section~\ref{sec:perf} evaluates the performance.  Section~\ref{sec:Conclusion} concludes the paper. 

\section{ Multi-Receiver Task-Oriented Communications} \label{sec:TOC}
\subsection{Task-oriented Communications}
Task-oriented or goal-oriented communications is a novel approach based on deep learning that presents an alternative to conventional wireless communications. Rather than relying on predefined protocols and algorithms, this paradigm leverages DNNs to learn and optimize the communication process for specific tasks. Deep learning models are trained on large amounts of data to learn the complex patterns and relationships within the communication process. This data-driven approach allows the adaptation and optimization of the end-to-end communications pipeline based on the specific task or goal at hand.  Instead of relying on separate modules for different tasks like channel coding and modulation, a deep learning model can learn to perform all these tasks jointly, enabling seamless integration and improved performance. In TOC, deep learning models can be trained to optimize the communication system for specific tasks or goals. For example, in a TOC system for computer vision applications, the model can learn to maximize the image classification accuracy under different channel conditions.

\begin{figure}[h]
\centering 
\begin{subfigure}[b]{0.49\textwidth}
\includegraphics[width=\columnwidth]{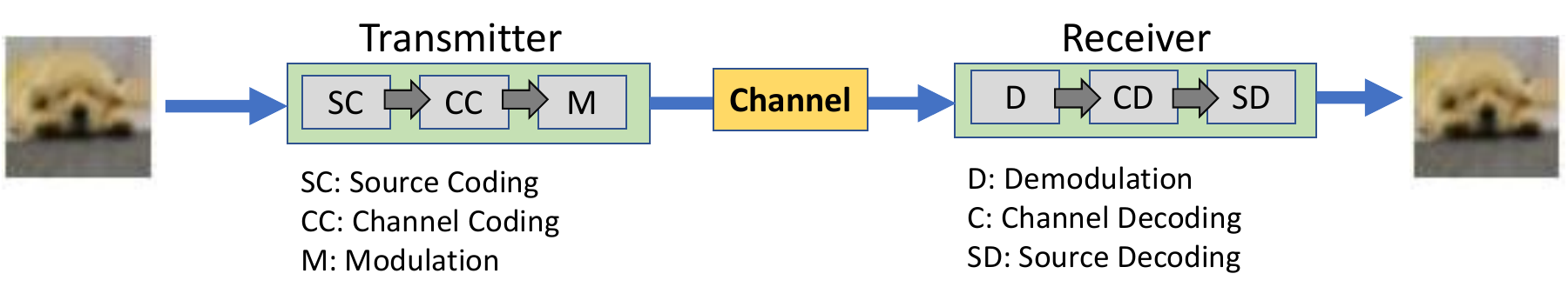}
\caption{Conventional communications.}
\label{fig:fig1}
\end{subfigure}
\begin{subfigure}[b]{0.49\textwidth}
\centering
\includegraphics[width=\columnwidth]{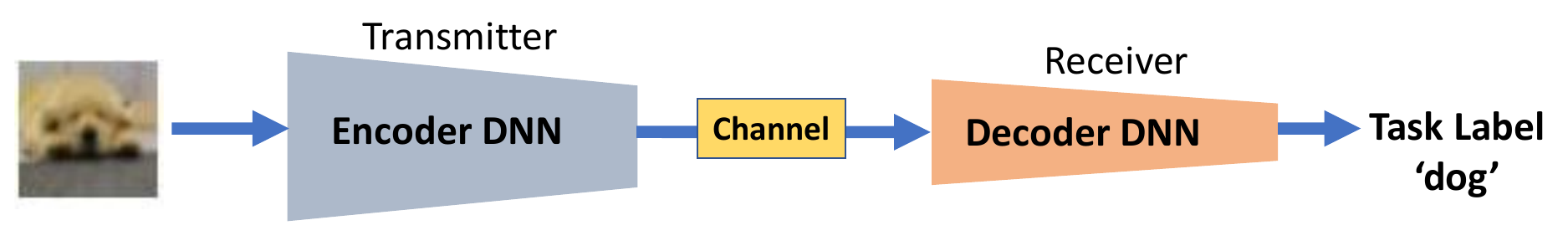}
\caption{Task-oriented communications.}
\label{fig:fig2}
\end{subfigure}
\caption{Conventional communications vs. TOC.}
\label{fig:figsub}
\end{figure}

We consider TOC driven by deep learning. The system model for STOC is shown in Fig.~\ref{fig:figsub} for the case of a single receiver. An encoder $\mathcal{E}$ and a decoder $\mathcal{D}$ are used to represent the transmitter and the receiver operations, respectively, and they are jointly trained by minimizing a loss $\mathcal{L}$ for the classification task $\mathcal{T}$. The encoder $\mathcal{E}$ performs source coding, channel coding, and modulation operations, transforming the input sample into modulated signals. The received signals at the receiver are classified by the decoder $\mathcal{D}$ to the labels of input data samples at the transmitter. Note that it is costly in terms of energy, channel use and delay to transmit the entire number of input samples separately. TOC uses the encoder $\mathcal{E}$ to reduce the dimension for the transmitted signals, thereby allowing more efficient allocation of resources. 

\subsection{MTOC}
\begin{figure}[t]
\centering
\includegraphics[width=\columnwidth]{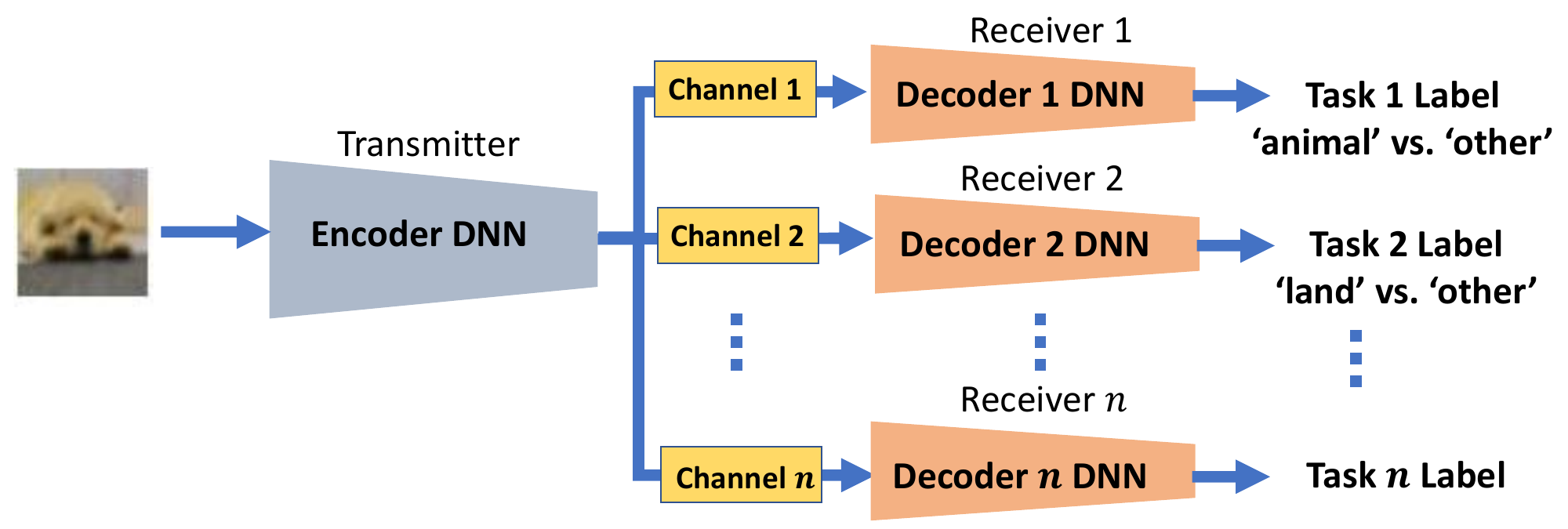}
 \caption{System diagram of MTOC.} 
 \label{fig:multitask}
\end{figure}

We consider one transmitter communicating with multiple ($n$) receivers as shown in Fig.~\ref{fig:multitask}. Each receiver $i$ has its own decoder $\mathcal{D}_i$ to perform a different classification task $\mathcal{T}_i$. The data samples such as images are the input to the encoder.  The common encoder $\mathcal{E}$ at the transmitter modulates the input data and broadcasts it in compressed form to decrease the number of channel uses. In other words, the size of the output of the encoder ($n_c$) is smaller than the size of the input sample such that the encoder $\mathcal{E}$ captures  latent features of low dimension that are transmitted with a reduced number of channel uses. Each receiver $i$ receives the same transmitted data experiencing a different channel and uses that as an input to its decoder $\mathcal{D}_i$. During training, the encoder $\mathcal{E}$ at the transmitter and the decoders $\left\{\mathcal{D}_i\right\}_{i=1}^n$ at all receivers are trained jointly as part of \emph{multi-task learning}. In case of STOC, the output of the encoder $\mathcal{E}$ needs to be transmitted to each receiver $i$ separately such that the encoder $\mathcal{E}$ and the decoder $\mathcal{D}_i$ of each receiver $i$ is trained together for each task $\mathcal{T}_i$ separately. In case of MTOC, the output of the encoder $\mathcal{E}$ is broadcast to all receivers at the same time, leading to a more efficient use of transmission opportunities by cutting transmission time $n$ times. 

Let $\boldsymbol{x}$ denote the input samples (e.g., images). The encoder's output at the transmitter is $\mathcal{E}(\boldsymbol{x})$ that is broadcast to $n$ receivers. Receiver $i$ receives signal $\boldsymbol{y}_i = \boldsymbol{h}_i \mathcal{E}(\boldsymbol{x})+ \boldsymbol{n}_i$, where $\boldsymbol{h}_i$ is the channel gain for receiver $i$ according to Rayleigh fading and $\boldsymbol{n}_i$ is the Gaussian noise at receiver $i$. This received signal is processed by decoder $\mathcal{D}_i$ such that the output label at receiver $i$ is $\hat{\mathcal{C}}_i (\boldsymbol{x}) = \mathcal{D}_i \left( \boldsymbol{h}_i \mathcal{E}(\boldsymbol{x})+ \boldsymbol{n}_i \right)$. Loss $\mathcal{L}_i$ measures the difference between the predicted label $\hat{\mathcal{C}}_i (\boldsymbol{x})$ and the true label $\mathcal{C} (\boldsymbol{x})$ for input sample $\boldsymbol{x}$.

\subsection{Multi-Task Learning}
We employ multi-task learning to jointly train the encoder at transmitter and the decoders at multiple receivers, each with its own task to complete. In multi-task learning, a model is trained to perform multiple (related) tasks simultaneously, instead of training separate models for each individual task. The idea of multi-task learning is that by jointly learning multiple tasks, the model can leverage the shared information across tasks for a more effective performance. Traditionally, in single-task learning, a model is trained to optimize a specific objective function for a particular task. On the other hand, multi-task learning allows the model to learn from multiple tasks simultaneously.

In multi-task learning, the model architecture is typically designed to have shared layers that capture common features across tasks, as well as task-specific layers that capture task-specific information. During training, the model is presented with examples from each task and learns to jointly optimize the objective functions of all the tasks. The objective functions can be weighted differently to reflect the relative importance of each task. Let $w_i \in [0,1]$ denote the weight for loss $\mathcal{L}_i$ for task $\mathcal{T}_i$ at receiver $i$. The combined loss for jointly training $\mathcal{E}$ and $\left\{\mathcal{D}_i\right\}_{i=1}^n$ is $\mathcal{L}_{\text{joint}} = \sum_{i=1}^n w_i \mathcal{L}_i$.    
 
The benefits of multi-task learning include: (i) \emph{Improved generalization}: By learning from multiple tasks, the model can capture common patterns and generalize better to unseen data. (ii) \emph{Regularization}: Training on multiple tasks can act as a form of regularization, preventing overfitting and improving the model's ability to handle noise and outliers. (iii) \emph{Data efficiency}: Multi-task learning can be particularly useful when there is limited data available for each individual task. By sharing information across tasks, the model can leverage the combined data to improve performance. (iv) \emph{Transfer learning}: Multi-task learning allows knowledge transfer between tasks, meaning that the model can benefit from the insights gained while solving one task to improve its performance on other related tasks. See examples of multi-task learning in \cite{mortaheb-spawc2022, mortaheb-ciss2023}.

\section{Datasets and Deep Neural Network Architectures} \label{sec:data}

We consider three types of image datasets:
\begin{itemize}
\item \textbf{MNIST:} The MNIST dataset is composed of grayscale images of handwritten digits \cite{MNIST}. The label of each data sample (image) is its digit (from $0$ to $9$). There are total of 10 classes and the set of classes is $C_{\text{MNIST}} = \{0,1,2,3,4,5,6,7,8,9\}$. Each sample is of $28\times28$ grayscale pixels and the value of each pixel is between $0$ and $255$. The dataset consists of 60,000 training samples and 10,000 test samples. 

\item \textbf{Fashion MNIST:} The Fashion-MNIST is composed of grayscale clothing images. There are total of 10 classes and the set of classes is $C_{\text{Fashion MNIST}} = \{$`T-shirt/top', `Trouser', `Pullover', `Dress', `Coat', `Sandal', `Shirt', `Sneaker', `Bag', `Ankle boot'$\}$.  Each sample is of $28\times28$ grayscale pixels and the value of each pixel is between $0$ and $255$. The dataset consists of 60,000 training samples and 10,000 test samples

\item \textbf{CIFAR-10:} The CIFAR-10 dataset consists of color images from 10 classes. The set of classes is $C_{\text{CIFAR-10}} = \{$ `Airplane',	`Automobile', `Bird', `Cat', `Deer', `Dog', `Frog', `Horse', `Ship', `Truck'$\}$ \cite{CIFAR}. Each data sample (image) is of $32 \times32 \times 3$ color (RGB) pixels and the value of a given pixel in each red, green and blue component is between $0$ and $255$. The dataset consists of 50,000 training samples and 10,000 test samples. 
\end{itemize}

Both MNIST and Fashion MNIST datasets can be effectively trained by either the FNN or CNN model. For the MNIST dataset, we consider the FNN, where each data sample is represented by a feature vector of size $28 \times 28 = 784$. For the Fashion MNIST dataset, we consider the CNN where each data sample is of size $28 \times 28 \times 1$. For the CIFAR-10 dataset, we consider the CNN as the model to train (the FNN is known to have poor performance for the CIFAR-10 dataset).

In each scenario, the corresponding feature is first normalized to a range of [0, 1]. Next, the normalized feature is used as input for the transmitter's encoder. The encoder reduces the dimension of each input sample to $n_c$, representing the number of channel uses required to transmit the modulated symbols from the transmitter's output (assuming one symbol can be transmitted per channel use). The encoded signal is then transmitted with $n_c$ channel uses over a Rayleigh fading channel with Gaussian noise added at the receiver. At each receiver, the received signal, also of dimension $n_c$, is passed as input to its decoder. The decoder's output provides the classification label. Notably, the input sample is not reconstructed at the receiver, which sets it apart from conventional communications.

We consider three data-DNN configurations: (a) the data is MNIST and the DNNs are FNN, (b) the data is Fashion MNIST and the DNNs are CNN, and (c) the data is CIFAR-10 and the DNNs are CNN. For these configurations,  the encoder and decoder architectures are outlined in Table~\ref{table:DNN}. The training process employs categorical cross-entropy as the loss function and utilizes the Adam optimizer. Rayleigh fading channel and Gaussian noise layers with the appropriate SNR are added between the encoder and each decoder. Numerical results are obtained using Python, and the models are trained in Keras with the TensorFlow backend. 

\begin{table}[h!]
 \captionsetup{justification=centering}
     \caption{Encoder-decoder architectures for task-oriented communications.}   
    \label{table:DNN}
	\begin{center}
	\footnotesize
        \begin{subtable}[h]{0.45\textwidth}
         \caption{Data: MNIST, Model: FNN.}
		\begin{tabular}{l|l|l}
			Network & Layer & Properties \\ \hline \hline
			Encoder & Input & size: 28$\times$28$\times$1 \\
            & Dense &  size: $256$, activation: ReLU \\
            & Dense &  size: $128$, activation: ReLU \\
			& Dense & size: $n_c$, activation: Linear \\ \hline
            Decoder & Input & size: $n_c$ \\ 
		for Task 1	& Dense & size: $n_c$, activation: ReLU \\
                & Dense & size: $\frac{n_c}{2}$, activation: ReLU \\
            & Dense & size: $2$, activation: Softmax
            \\ \hline
            Decoder & Input & size: $n_c$ \\ 
		for Task 2	& Dense & size: $n_c$, activation: ReLU \\
                & Dense & size: $\frac{n_c}{2}$, activation: ReLU \\
            & Dense & size: $2$, activation: Softmax \\
		\end{tabular}
        \end{subtable}
        
\vspace{0.2cm}

        \begin{subtable}[h]{0.45\textwidth}
            \caption{Data: Fashion MNIST, Model: CNN.}
		\begin{tabular}{l|l|l}
			Network & Layer & Properties \\ \hline \hline
			Encoder & Input & size: 28$\times$28$\times$1 \\
& Conv2D & filter size: 32, kernel size: (3,3) \\ & & activation: ReLU \\
& MaxPooling2D & pool size: (2,2) \\
& Conv2D & filter size: 32, kernel size: (3,3) \\ & & activation: ReLU \\
& MaxPooling2D & pool size: (2,2) \\
& Flatten & -- \\
& Dropout & dropout rate: 0.5 \\
& Dense & size: 128, activation: ReLU \\ 
& Dense & size: $n_c$, activation: Linear \\  \hline
            Decoder & Input & size: $n_c$ \\ 
		for Task 1	& Dense & size: $n_c$, activation: ReLU \\
                & Dense & size: $\frac{n_c}{2}$, activation: ReLU \\
            & Dense & size: $2$, activation: Softmax
            \\ \hline
            Decoder & Input & size: $n_c$ \\ 
		for Task 2	& Dense & size: $n_c$, activation: ReLU \\
                & Dense & size: $\frac{n_c}{2}$, activation: ReLU \\
            & Dense & size: $2$, activation: Softmax \\
		\end{tabular}
        \end{subtable}
        
\vspace{0.2cm}

        \begin{subtable}[h]{0.45\textwidth}
        \caption{Data: CIFAR-10, Model: CNN.}
		\begin{tabular}{l|l|l}
			Network & Layer & Properties \\ \hline \hline
			Encoder & Input & size: 32$\times$32$\times$3 \\
& Conv2D & filter size: 8, kernel size: (3,3) \\ & & activation: ReLU \\
& Conv2D & filter size: 4, kernel size: (3,3) \\ & & activation: ReLU \\
& MaxPooling2D & pool size: (2,2) \\
& Dropout & dropout rate: 0.1 \\
& Conv2D & filter size: 4, kernel size: (3,3) \\ & & activation: ReLU \\
& MaxPooling2D & pool size: (2,2) \\
& Dropout & dropout rate: 0.1 \\
& Flatten & -- \\
& Dense & size: 128, activation: ReLU \\
& Dense & size: $n_c$, activation: Linear \\ \hline
            Decoder & Input & size: $n_c$ \\ 
		for Task 1	& Dense & size: $n_c$, activation: ReLU \\
                & Dense & size: $\frac{n_c}{2}$, activation: ReLU \\
            & Dense & size: $2$, activation: Softmax
            \\ \hline
            Decoder & Input & size: $n_c$ \\ 
		for Task 2	& Dense & size: $n_c$, activation: ReLU \\
                & Dense & size: $\frac{n_c}{2}$, activation: ReLU \\
            & Dense & size: $2$, activation: Softmax \\
		\end{tabular}
        \end{subtable}
        
	\end{center}
 \vspace{-0.4cm}
\end{table}

\section{Performance Evaluation} \label{sec:perf}
We start with two receivers and consider two different tasks for each data type. For the MNIST data, task 1 is to classify images to two subsets, even digits and odd digits, and task 2 is to classify images to two subsets, digits smaller than 5 and larger than or equal to 5. For the Fashion MNIST data, task 1 is to classify images to two subsets, dress items (`T-shirt/top', `Trouser', `Pullover', `Dress', `Coat', `Shirt') and others, and task 2 is to classify images to two subsets, formal clothing (`Trouser', `Dress', 'Sandal, Shirt, `Bag') and others. For the CIFAR-10 data, task 1 is to classify images to 
animals (`bird', `cat', `deer', `dog', `frog', `horse') and others, and task 2 is to classify images to  small ground entities (`automobile', `cat', `deer', `dog', `horse') and others.

\begin{figure}[h]
	\centering
	\includegraphics[width=0.81\columnwidth]{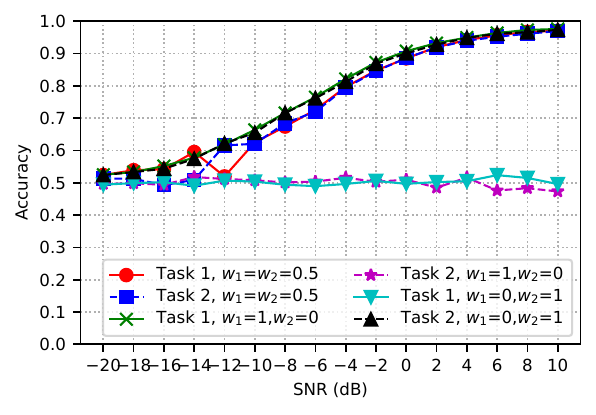}
	\caption{Task accuracy vs. SNR for MNIST+FNN.}
	\label{fig:MNISTvarydB}
 \vspace{0.2cm}
% \end{figure}

% \begin{figure}[h]
	\centering
	\includegraphics[width=0.81\columnwidth]{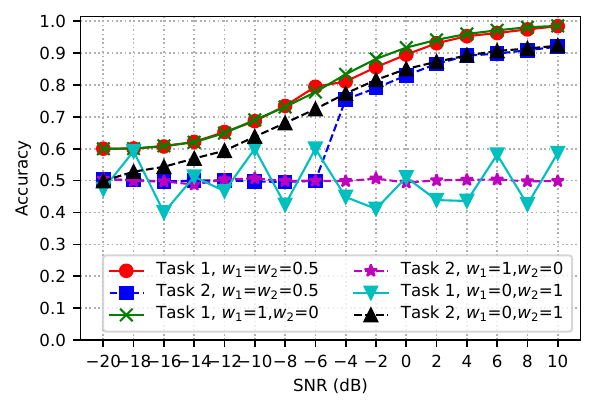}
	\caption{Task accuracy vs. SNR for Fashion MNIST+CNN.}
	\label{fig:FashionvarydB}
  \vspace{0.2cm}
% \end{figure}

% \begin{figure}[h]
	\centering
	\includegraphics[width=0.81\columnwidth]{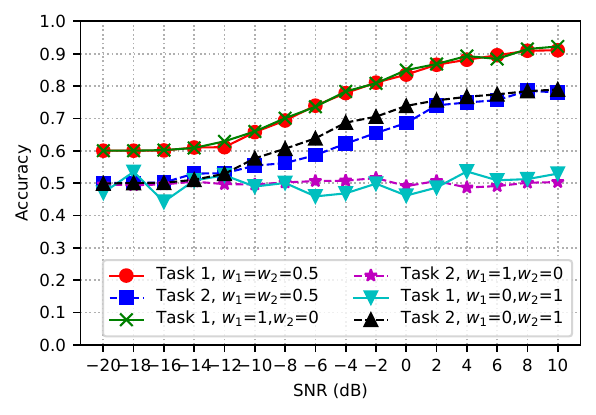}
	\caption{Task accuracy vs. SNR for CIFAR-10+CNN.}
	\label{fig:CIFARvarydB}
\end{figure}

\begin{figure}[h]
	\centering
	\includegraphics[width=0.81\columnwidth]{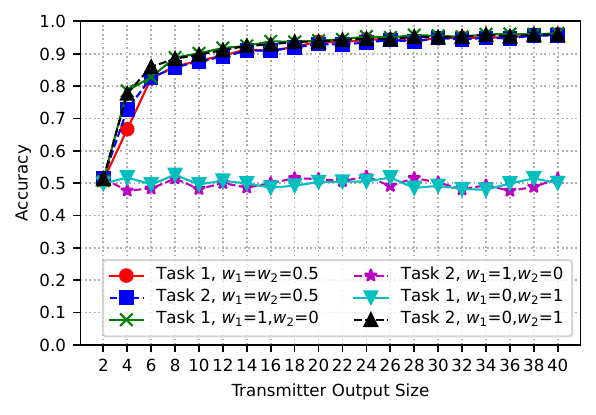}
	\caption{Task accuracy vs. transmitter output size for MNIST+FNN.}
	\label{fig:MNISTvarync}
  \vspace{0.2cm}
% \end{figure}

% \begin{figure}[h]
	\centering
	\includegraphics[width=0.81\columnwidth]{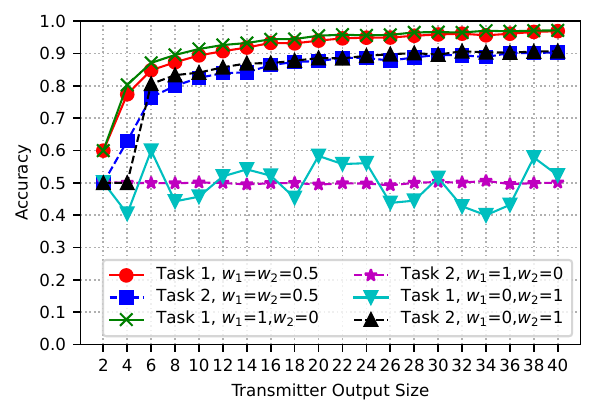}
	\caption{Task accuracy vs. transmitter output size for Fashion MNIST+CNN.}
	\label{fig:Fashionvarync}
 \vspace{0.2cm}
% \end{figure}

% \begin{figure}[h]
	\centering
	\includegraphics[width=0.81\columnwidth]{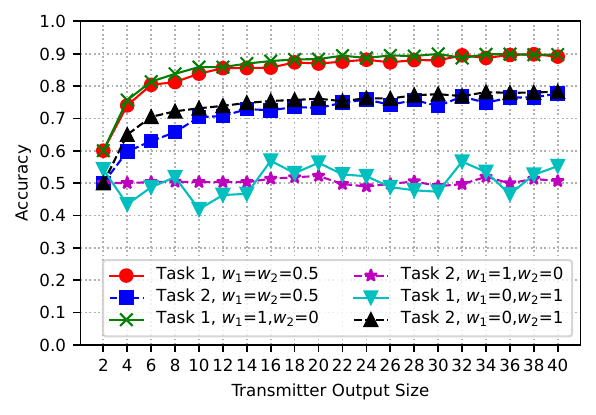}
	\caption{Task accuracy vs. transmitter output size for CIFAR-10+CNN.}
	\label{fig:CIFARvarync}
\end{figure}

\begin{figure}[h]
	\centering
	\includegraphics[width=0.81\columnwidth]{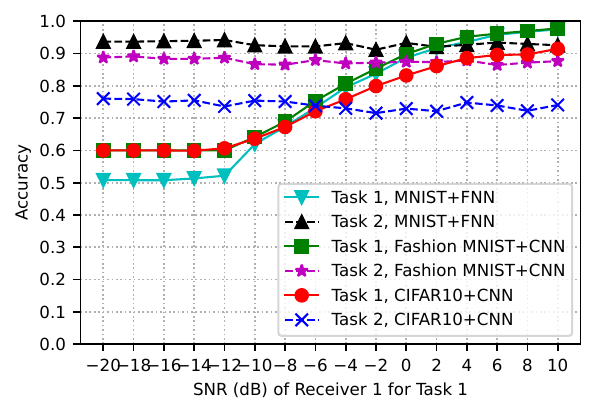}
	\caption{Task accuracy vs. SNR for receiver 1 (SNR for receiver 2 is 3dB).}
	\label{fig:varydB1}
  \vspace{0.2cm}
% \end{figure}

% \begin{figure}[h]
	\centering
	\includegraphics[width=0.81\columnwidth]{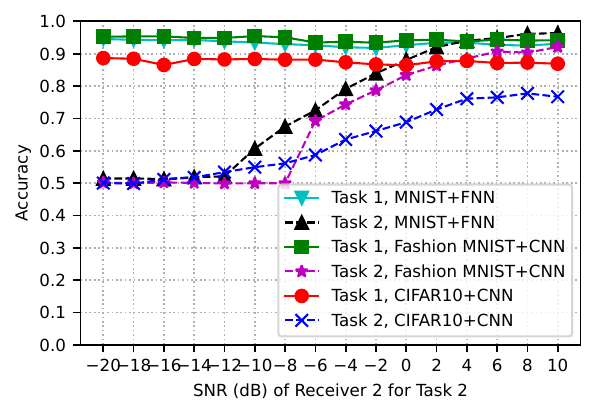}
	\caption{Task accuracy vs. SNR for receiver 2 (SNR for receiver 1 is 3dB).}
	\label{fig:varydB2}
  \vspace{0.2cm}
% \end{figure}

% \begin{figure}[h]
	\centering
	\includegraphics[width=0.81\columnwidth]{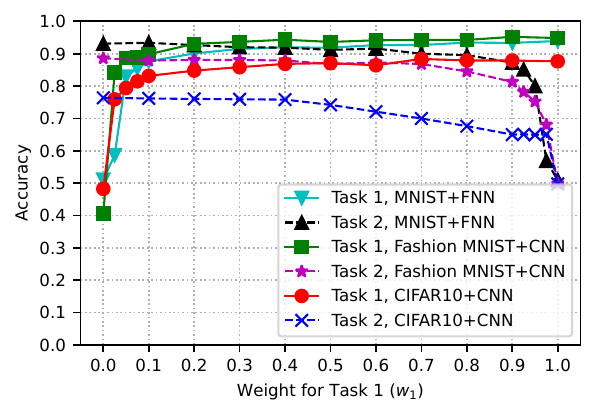}
	\caption{Task accuracy vs. weight $w_1$ for task 1 (where $w_2=1-w_1$).}
	\label{fig:varygamma}
    \vspace{-0.3cm}
\end{figure}

Default values for SNR and $n_c$ are 3dB and 20, respectively, assuming common values for each receiver. Figs. \ref{fig:MNISTvarydB}, \ref{fig:FashionvarydB}, and \ref{fig:CIFARvarydB} show the accuracy for both tasks as a function of the SNR for the three data-DNN configurations (a), (b), and (c), respectively. Figs. \ref{fig:MNISTvarync}, \ref{fig:Fashionvarync}, and \ref{fig:CIFARvarync} show the accuracy for both tasks as a function of the transmitter's output size, $n_c$, for the three data-DNN configurations (a), (b), and (c), respectively. Different weights, $w_i$, $i=1,2$ are applied to the loss values during training that include cases $(w_1=1, w_2=0)$, $(w_1=0, w_2=1)$, and $(w_1=w_2=0.5)$. For receiver $i$ when the weight $w_i$ is set to $1$ during training (this corresponds to the case of STOC for task $\mathcal{T}_i$ only), it is trained well, and the accuracy of task $\mathcal{T}_i$ increases with increasing SNR and $n_c$. On the other hand, training is not successful and decoder $\mathcal{D}_i$ makes random decisions when the weight for  task $\mathcal{T}_i$ is set to $0$. Both tasks $\mathcal{T}_1$ and $\mathcal{T}_2$ at two receivers achieve high accuracy when $w_i=0.5$, $i=1,2$. Hence, partitioning weights among tasks as needed in MTOC can achieve high accuracy for both tasks compared to STOC. Note that serving both tasks separately with STOC would require doubling the transmitter's output size. Therefore, MTOC can achieve high accuracy with fewer transmissions. These results continue to hold when we have asymmetric channel conditions for both receivers, as shown in Fig.~\ref{fig:varydB1} and Fig.~\ref{fig:varydB2}, where we change the SNR for one receiver while keeping the SNR for the other receiver fixed.

We have a closer look at the effects of weights in Fig.~\ref{fig:varygamma}. Even a small non-zero weight assigned to a task leads to high accuracy for that task, thereby motivating the use of MTOC.

\begin{figure}[h]
	\centering
	\includegraphics[width=0.81\columnwidth]{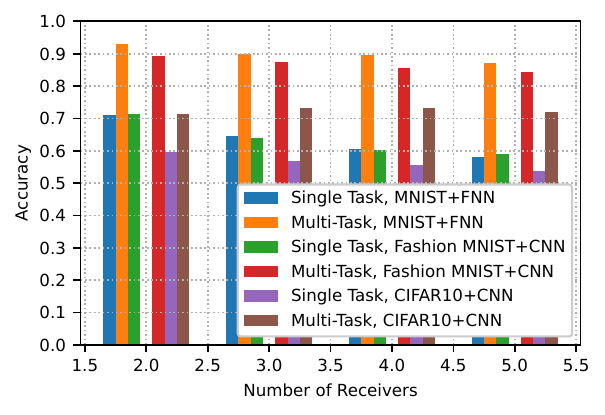}
	\caption{Task accuracy vs.  number of receivers.}
	\label{fig:varynumrec}
    \vspace*{-0.4cm}
\end{figure}

Next, we increase the number of receivers $n$. Fig.~\ref{fig:varynumrec} shows the task accuracy as a function of $n$, when each receiver completes its individual task. In this scenario, the task for receiver $i$, where $1 \leq i \leq n $, is to classify images to the subset of classes $C_j$, $j = i, ..., (i+4) \text{ mod } 10$ and the rest of classes, where $C_j$ is the $j$th class in the class set ($C_{\text{MNIST}}$ for the MNIST dataset, $C_{\text{Fashion MNIST}}$ for the Fashion MNIST dataset, and $C_{\text{CIFAR-10}}$ for the CIFAR-10 dataset). As $n$ increases, the accuracy averaged over all receivers drops when we consider STOC targeting the task at one receiver only, whereas MTOC maintains the high task accuracy for all receivers.

% % This is for Semantic
% \begin{figure}[t]
% 	\centering
% 	\includegraphics[width=0.81\columnwidth]{figures/semanticvarydB.pdf}
% 	\caption{Semantic performance as a function of the SNR.}
% 	\label{fig:semanticvarydB}
% \end{figure}

\section{Conclusion} \label{sec:Conclusion}
We have studied TOC, where a transmitter establishes communications with multiple receivers, each assigned a distinct task involving a shared dataset, such as images sourced from the transmitter. We have presented an innovative multi-task deep learning approach that entails training a common encoder at the transmitter, while equipping individual decoders at each receiver. The overarching goal is to jointly optimize multiple tasks and communication, all while ensuring resource allocation efficiency at the edge of 6G networks. The aim is to enable the system to adapt seamlessly to varying channel conditions and task characteristics, while effectively accomplishing task-specific objectives and minimizing transmission overhead. Central to this approach is the joint training of the encoder and decoders through multi-task learning, facilitating the assimilation of shared information across tasks and optimizing the communication process accordingly. Leveraging the broadcast property of wireless communications, the MTOC approach significantly diminishes the number of transmissions required to complete tasks at different receivers. We have conducted comprehensive performance evaluation on MNIST, Fashion MNIST, and CIFAR-10 datasets, focusing on various image classification tasks. The findings demonstrate the substantial efficacy of MTOC, outperforming STOC in terms of classification accuracy and resource utilization.

\bibliographystyle{IEEEtran}
\bibliography{references}

\end{document}